\documentclass[10pt,letter,prl,twocolumn]{article}

\usepackage{ifthen} 
\newboolean{pdflatex} 
\setboolean{pdflatex}{true} 

\newboolean{articletitles}
\setboolean{articletitles}{true} 

\newboolean{uprightparticles}
\setboolean{uprightparticles}{false} 

\usepackage{microtype}
\usepackage{lineno}   
\usepackage{xspace} 

\usepackage{graphicx}  
\usepackage{subfigure}
\usepackage{color}
\usepackage{colortbl}
\graphicspath{{./}} 

\usepackage{amsmath} 
\usepackage{amssymb}
\usepackage{amsfonts}
\usepackage{upgreek} 

\newcommand*\patchAmsMathEnvironmentForLineno[1]{%
\expandafter\let\csname old#1\expandafter\endcsname\csname #1\endcsname
\expandafter\let\csname oldend#1\expandafter\endcsname\csname
end#1\endcsname
 \renewenvironment{#1}%
   {\linenomath\csname old#1\endcsname}%
   {\csname oldend#1\endcsname\endlinenomath}%
}
\newcommand*\patchBothAmsMathEnvironmentsForLineno[1]{%
  \patchAmsMathEnvironmentForLineno{#1}%
  \patchAmsMathEnvironmentForLineno{#1*}%
}
\AtBeginDocument{%
\patchBothAmsMathEnvironmentsForLineno{equation}%
\patchBothAmsMathEnvironmentsForLineno{align}%
\patchBothAmsMathEnvironmentsForLineno{flalign}%
\patchBothAmsMathEnvironmentsForLineno{alignat}%
\patchBothAmsMathEnvironmentsForLineno{gather}%
\patchBothAmsMathEnvironmentsForLineno{multline}%
}

\usepackage{hyperref}    
\usepackage[all]{hypcap} 

\def\pythia     {\mbox{\textsc{Pythia}}\xspace}
\def\evtgen     {\mbox{\textsc{EvtGen}}\xspace}
\def\geant      {\mbox{\textsc{Geant4}}\xspace}
\def\gauss      {\mbox{\textsc{Gauss}}\xspace}
\def\lhcb {\mbox{LHCb}\xspace}
\newcommand{\mevcc}{\ensuremath{{\mathrm{\,Me\kern -0.1em V\!/}c^2}}\xspace}
\newcommand{\gevcc}{\ensuremath{{\mathrm{\,Ge\kern -0.1em V\!/}c^2}}\xspace}
\newcommand{\tev}{\ensuremath{\mathrm{\,Te\kern -0.1em V}}\xspace}
\newcommand{\gevc}{\ensuremath{{\mathrm{\,Ge\kern -0.1em V\!/}c}}\xspace}
\def\mum  {\ensuremath{\,\upmu\rm m}\xspace}
\def\PJ      {\ensuremath{\mathrm{J}}\xspace}                 
\def\Ppsi        {\ensuremath{\psi}\xspace}                 

\def\BF         {{\ensuremath{\cal B}\xspace}}
\def\invfb   {\ensuremath{\mbox{\,fb}^{-1}}\xspace}
\def\BR         {\BF}
\def\Kbar  {\kern 0.2em\overline{\kern -0.2em \PK}{}\xspace}
\def\PK      {\ensuremath{\mathrm{K}}\xspace}                 
\def\jpsi     {\ensuremath{{\PJ\mskip -3mu/\mskip -2mu\Ppsi\mskip 2mu}}\xspace}
\def\psitwos  {\ensuremath{\Ppsi{(2S)}}\xspace}
                 
\def\mup        {\ensuremath{\mu^+}\xspace}
\def\mun        {\ensuremath{\mu^-}\xspace} 
\def\Ps      {\ensuremath{\mathrm{s}}\xspace}                 
\def\squark    {\ensuremath{\Ps}\xspace}

\def\Bu      {\ensuremath{\B^+}\xspace}
\def\Bsb     {\ensuremath{\Bbar^0_\squark}\xspace}
\def\Bbar    {\kern 0.18em\overline{\kern -0.18em \PB}{}\xspace}
\def\PB      {\ensuremath{B}\xspace}

\def\B       {\ensuremath{\PB}\xspace}
\def\Bd      {\ensuremath{\B^0}\xspace}
\def\Bs      {\ensuremath{\B^0_\squark}\xspace}
\def\Bp      {\ensuremath{\Bu}\xspace}

\newcommand{\BqSP}{\mbox{\ensuremath{B^0_{(s)}\rightarrow SP }}\xspace}
\newcommand{\BsSP}{\ensuremath{\Bs\to S  P }\xspace}
\newcommand{\BdSP}{\ensuremath{\Bd\to S  P }\xspace}
\newcommand{\BdJpsiKpi}{\mbox{\ensuremath{\Bd \to J/\psi K^+\pi^-}}\xspace}
\newcommand{\Bq}{\mbox{\ensuremath{B^0_{(s)} }}\xspace} 

\newcommand{\KstKpi}{\ensuremath{K^{*0}\to K^+\pi^-}\xspace}
\newcommand{\Kst}{\ensuremath{K^{*0}}\xspace}
\newcommand{\EBqSP}{\mbox{\ensuremath{\epsilon_{B^0_{(s)}\rightarrow SP }}}\xspace}
\newcommand{\EBsmmmm}{\ensuremath{\epsilon_{\Bs\to\mu^+\mu^-\mu^+\mu^-}}\xspace}

\newcommand{\EBqmmmm}{\ensuremath{\epsilon_{\B^0_{(s)}\to\mu^+\mu^-\mu^+\mu^-}}\xspace}
\newcommand{\NBqmmmm}{\ensuremath{N_{\B^0_{(s)}\to\mu^+\mu^-\mu^+\mu^-}}\xspace}
\newcommand{\Bsmmmm}{\mbox{\ensuremath{\Bs\to\mu^+\mu^-\mu^+\mu^-}}\xspace}
\newcommand{\Bdmmmm}{\mbox{\ensuremath{\Bd\to\mu^+\mu^-\mu^+\mu^-}}\xspace}
\newcommand{\Bqmmmm}{\mbox{\ensuremath{B^0_{(s)}\to \mup\mun\mup\mun}}\xspace}
\newcommand{\BsJphi}{\mbox{\ensuremath{\Bs \to J/\psi\phi }}\xspace}

\def\BsJPsimmPhimm {\ensuremath{\Bs \to  \jpsi \, (\to \mu^+ \mu^-) \, \phi \, (\to \mu^+ \mu^-)}\xspace}

\newcommand{\BdJpsiKst}{\ensuremath{\Bd \to J/\psi K^{*0}}\xspace}
\newcommand{\BsJpsiKst}{\ensuremath{\Bs\to J/\psi \Kbar^{*0}}\xspace}
\newcommand{\BsbJpsiKst}{\ensuremath{\Bsb\to J/\psi K^{*0}}\xspace} 
\newcommand{\BdJpsimumuKstKpi}{\ensuremath{\Bd\to J/\psi\,(\to\mu^+\mu^-)\,K^{*0}\,(\to K^+\pi^-)}\xspace}

\newcommand{\Jpsimumu}{\mbox{\ensuremath{J/\psi\to \mu^+\mu^-}}\xspace}

\newcommand{\Jpsi}{\ensuremath{J/\psi}\xspace}

\newcommand\Tstrut{\rule{0pt}{2.6ex}}

\newcommand\Bstrut{\rule[-1.2ex]{0pt}{0pt}}

\usepackage{mciteplus} 
\usepackage{tabularx}
\begin{document}

\renewcommand{\thefootnote}{\fnsymbol{footnote}}
\setcounter{footnote}{1}

\onecolumn

\begin{titlepage}
\pagenumbering{roman} 

\vspace*{-1.5cm}
\centerline{\large EUROPEAN ORGANIZATION FOR NUCLEAR RESEARCH (CERN)}
\vspace*{1.5cm}
\hspace*{-0.5cm}
\begin{tabular*}{\linewidth}{lc@{\extracolsep{\fill}}r}
\\
 & & CERN-PH-EP-2013-031 \\  
 & & LHCb-PAPER-2012-049 \\  
 & & March 5, 2013 \\ 
 & & \\
\end{tabular*}

\vspace*{4.0cm}

{\bf\boldmath\huge
\begin{center}
  Search for Rare $B^0_{(s)}\rightarrow \mu^+ \mu^- \mu^+ \mu^-$ Decays
\end{center}
}

\vspace*{2.0cm}

\begin{center}
The LHCb Collaboration\footnote{Authors are listed on the following pages.}
\end{center}

\vspace{\fill}

\begin{abstract}
  \noindent
A search for the decays $B^0_{s}\rightarrow \mu^+ \mu^- \mu^+ \mu^-$ and $B^0 \rightarrow \mu^+ \mu^- \mu^+ \mu^-$ is performed using data, corresponding to an integrated luminosity of 1.0\ensuremath{\mbox{\,fb}^{-1}}\xspace, collected with the LHCb detector in 2011. The number of candidates observed is consistent with the expected background and, assuming phase-space models of the decays, limits on the branching fractions are set: \mbox{${\ensuremath{\cal B}\xspace}(B^0_{s}\rightarrow \mu^+ \mu^- \mu^+ \mu^-) < 1.6 \ (1.2) \times 10^{-8}$}  and \mbox{${\ensuremath{\cal B}\xspace}(B^0 \rightarrow \mu^+ \mu^- \mu^+ \mu^-)< 6.6 \ (5.3) \times 10^{-9}$} at 95\,\% (90\,\%) confidence level. In addition, limits are set in the context of a supersymmetric model which allows for the $B^0_{(s)}$ meson to decay into a scalar ($S$) and pseudoscalar particle ($P$), where $S$ and $P$ have masses of 2.5\,\ensuremath{{\mathrm{\,Ge\kern -0.1em V\!/}c}}\xspace and 214.3\,\ensuremath{{\mathrm{\,Me\kern -0.1em V\!/}c}}\xspace, 
respectively, both resonances decay into $\mu^+\mu^-$. The branching fraction limits for these decays are \mbox{${\ensuremath{\cal B}\xspace}(\ensuremath{B^0_{s}\rightarrow SP }\xspace) < 1.6 \ (1.2) \times 10^{-8}$} and \mbox{${\ensuremath{\cal B}\xspace}(\ensuremath{B^0\rightarrow SP }\xspace)< 6.3 \ (5.1) \times 10^{-9}$} at 95\,\% (90\,\%) confidence level. 

\end{abstract}

\vspace*{2.0cm}

\begin{center}
  Submitted to Physical Review Letters
\end{center}

\vspace{\fill}

{\footnotesize
\centerline{\copyright~CERN on behalf of the \lhcb collaboration, license \href{http://creativecommons.org/licenses/by/3.0/}{CC-BY-3.0}.}}
\vspace*{2mm}

\end{titlepage}


\newpage
\setcounter{page}{2}
\mbox{~}

\centerline{\large\bf LHCb collaboration}
\begin{flushleft}
\small
R.~Aaij$^{38}$, 
C.~Abellan~Beteta$^{33,n}$, 
A.~Adametz$^{11}$, 
B.~Adeva$^{34}$, 
M.~Adinolfi$^{43}$, 
C.~Adrover$^{6}$, 
A.~Affolder$^{49}$, 
Z.~Ajaltouni$^{5}$, 
J.~Albrecht$^{9}$, 
F.~Alessio$^{35}$
% ...

{\footnotesize \it
$ ^{1}$Centro Brasileiro de Pesquisas F\'{i}sicas (CBPF), Rio de Janeiro, Brazil\\
% ...
$ ^{5}$Clermont Universit\'{e}, Universit\'{e} Blaise Pascal, CNRS/IN2P3, LPC, Clermont-Ferrand, France\\
$ ^{6}$CPPM, Aix-Marseille Universit\'{e}, CNRS/IN2P3, Marseille, France\\
% ...
$ ^{9}$Fakult\"{a}t Physik, Technische Universit\"{a}t Dortmund, Dortmund, Germany\\
% ...
$ ^{11}$Physikalisches Institut, Ruprecht-Karls-Universit\"{a}t Heidelberg, Heidelberg, Germany\\
% ...
$ ^{33}$Universitat de Barcelona, Barcelona, Spain\\
$ ^{34}$Universidad de Santiago de Compostela, Santiago de Compostela, Spain\\
$ ^{35}$European Organization for Nuclear Research (CERN), Geneva, Switzerland\\
% ...
$ ^{38}$Nikhef National Institute for Subatomic Physics, Amsterdam, The Netherlands\\
% ...
$ ^{43}$H.H. Wills Physics Laboratory, University of Bristol, Bristol, United Kingdom\\
% ...
$ ^{49}$Oliver Lodge Laboratory, University of Liverpool, Liverpool, United Kingdom\\
% ...
$ ^{n}$Also at LIFAELS, La Salle, Universitat Ramon Llull, Barcelona, Spain\\
% ...
}
\end{flushleft}
\cleardoublepage

\twocolumn

\renewcommand{\thefootnote}{\arabic{footnote}}
\setcounter{footnote}{0}



\pagestyle{plain} 
\setcounter{page}{1}
\pagenumbering{arabic}


%


  \begin{figure*}
 \centering
   \subfigure[]{\includegraphics[height = 0.18\textheight]{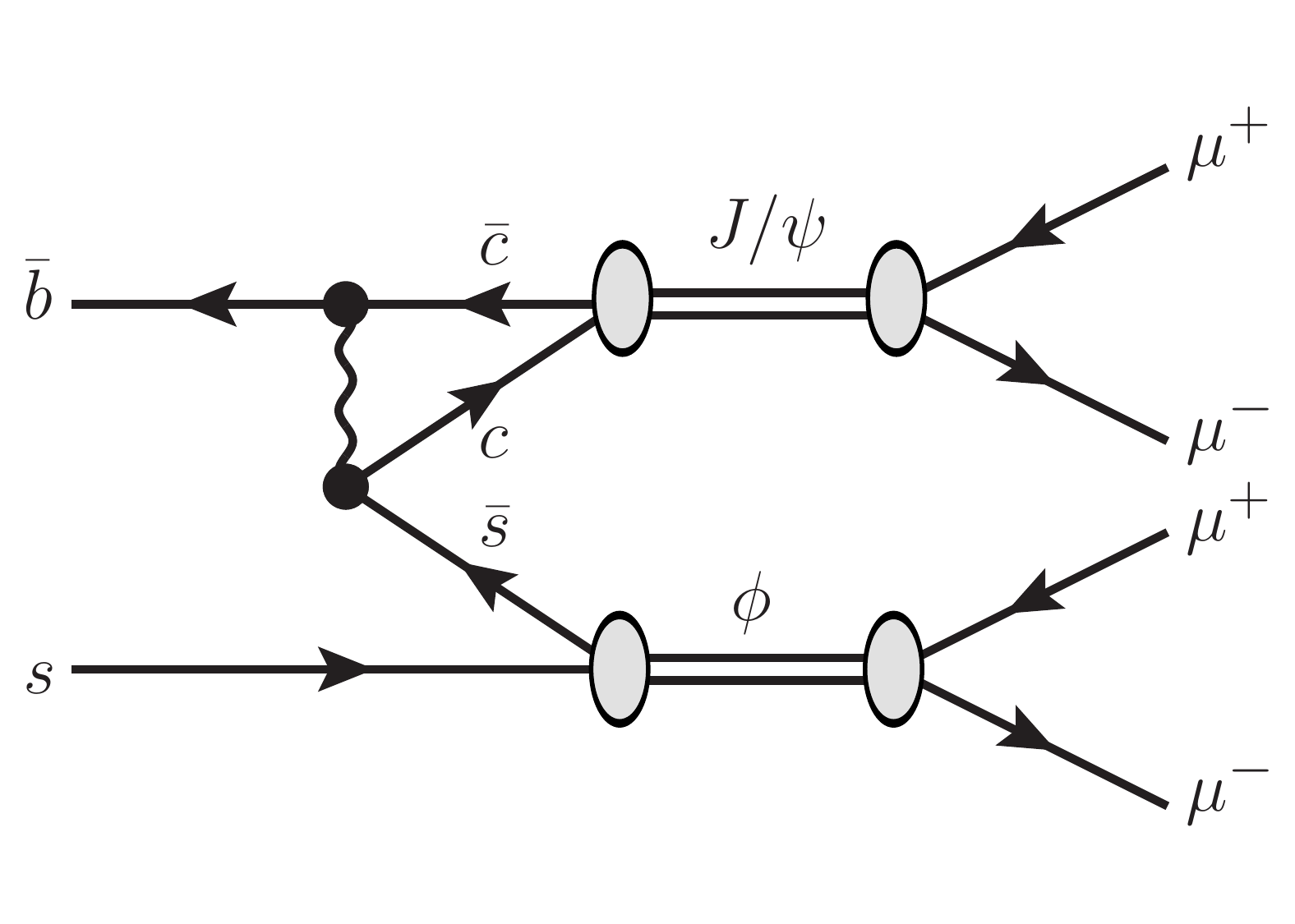}}
      \subfigure[]{\includegraphics[height = 0.17\textheight]{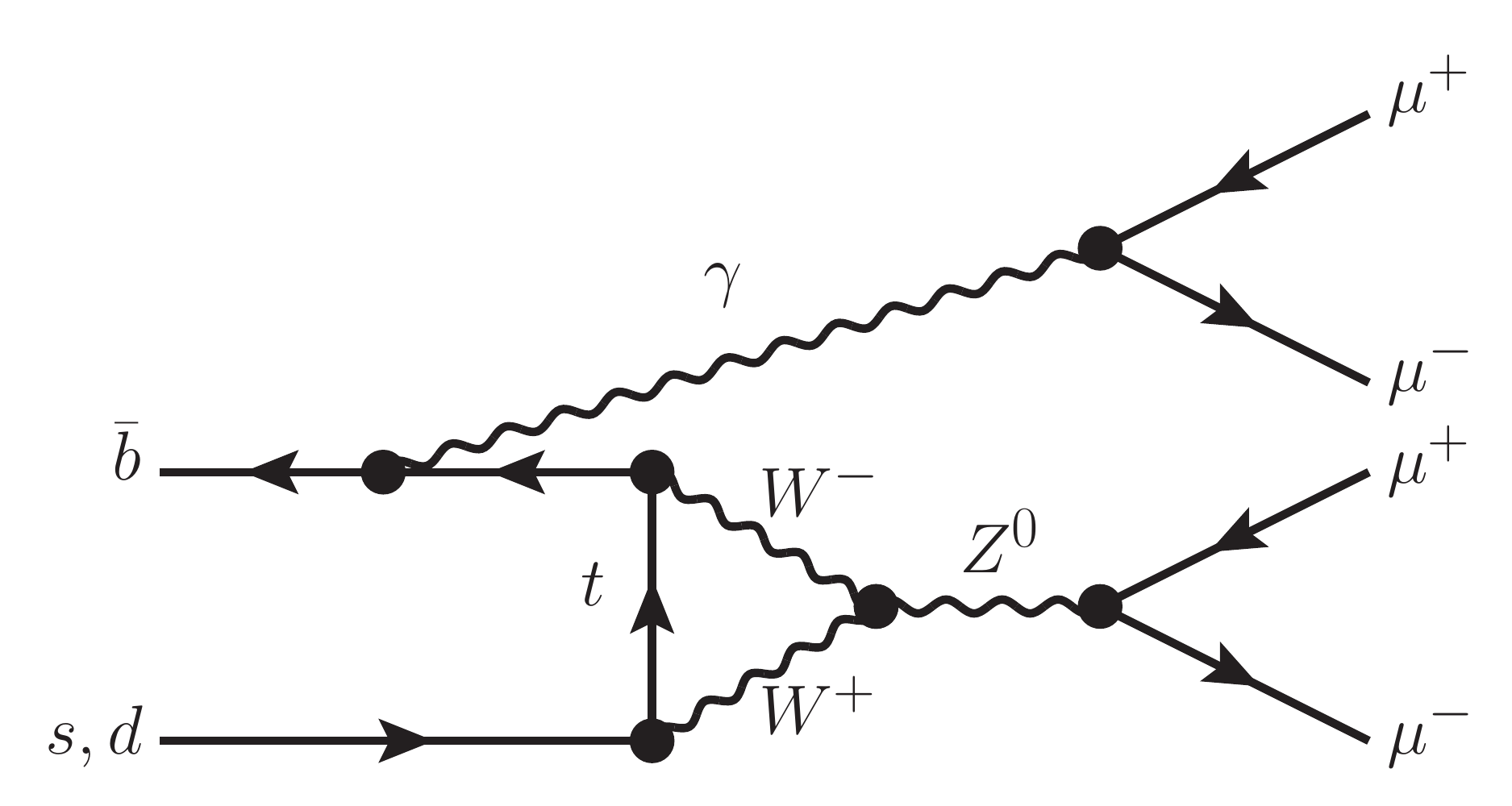}}
      \subfigure[]{\includegraphics[height = 0.17\textheight]{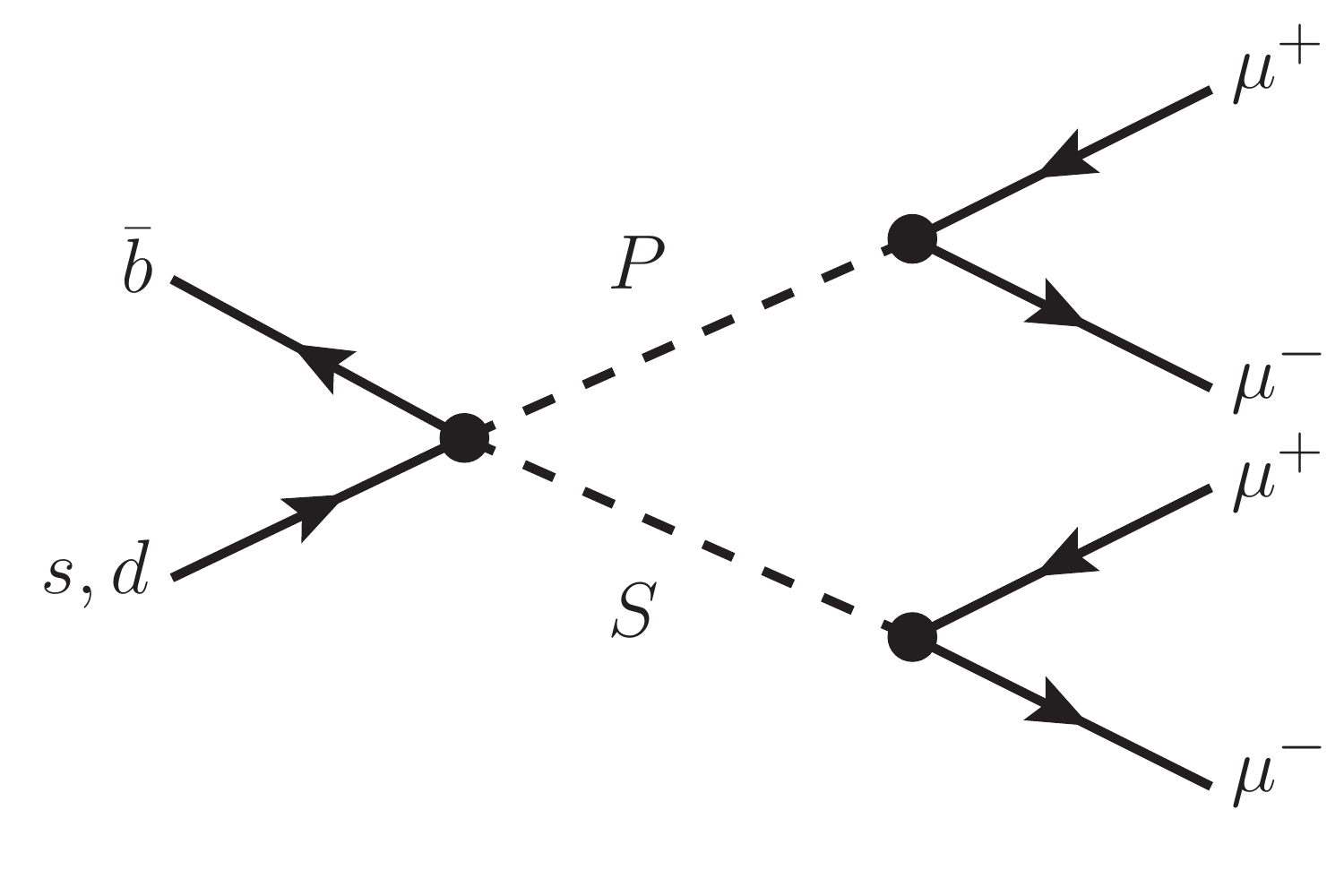}}

   \caption{Feynman diagrams for the \Bsmmmm and \Bdmmmm decays, via (a) the resonant \mbox{\BsJphi} SM channel, (b) the nonresonant SM channel and (c) the supersymmetric channel. The latter is propagated by scalar $S$ and pseudoscalar $P$ sgoldstino sfermions.}
\label{fig:feyns}
\end{figure*} 

The decays \Bqmmmm are strongly suppressed in the standard model (SM). Significant enhancements in the branching fractions can occur in beyond the SM theories~\cite{Demidov,axion}. For example, in minimal supersymmetric models (MSSM) the decay is mediated via the new scalar $S$, and pseudoscalar $P$ sgoldstino particles, which both decay into $\mup\mun$. The sgoldstinos can couple to SM particles via type-I couplings, where one sgoldstino couples to two SM fermions and via type-II couplings, where both $S$ and $P$ couple to two SM fermions in a four-prong vertex~\cite{Demidov}. Such searches are also of particular interest since the HyperCP Collaboration found evidence of the decay \mbox{$\Sigma^+ \rightarrow p \mup \mun$}, which is consistent with the decay \mbox{$\Sigma^+\rightarrow p P$} and \mbox{$P\rightarrow \mup\mun$} with a $P$ particle mass of $214.3\pm 0.5$\,\mevcc~\cite{Park:2005eka}. The inclusion of charge conjugated processes is implied throughout this Letter. The Belle Collaboration has
reported a search for 
a $P$ particle in \mbox{$\Bd \to V P(\to \mup \mun)$} decays, where $V$ is either a $\Kst(892)$ or a $\rho^0(770)$ meson~\cite{Hyun:2010an}. The corresponding upper limits constrain the type-I couplings of 
sgoldstinos to the SM particles. The search presented here is sensitive to the decays \BqSP in which both the $S$ and $P$ particle decay vertices are not significantly displaced from the \Bq decay vertex and the $P$ particle has a similar mass to that reported by the HyperCP collaboration. Such final states probe type-II couplings, which might dominate for general low energy supersymmetry breaking models and have not been probed by previous experiments. The present search is therefore complementary to the Belle study. Moreover, the four-muon final state is essentially background free and therefore ideally suited to search for new physics signatures. The present search is also sensitive to \Bqmmmm decays which are not propagated through intermediate resonant structures. 

The dominant SM decay of a $B$ meson into a four-muon final state is \mbox{$\BsJphi(1020)$}, where both the $J/\psi$ and the $\phi$ mesons decay into two muons \mbox{[Fig.~\ref{fig:feyns}\,(a)]}. In this Letter, this is referred to as the resonant decay mode. The branching fraction for \mbox{\BsJPsimmPhimm} is calculated as the product of \BF(\BsJphi), \BF(\Jpsimumu) and \BF(\mbox{$\phi \rightarrow \mu^+ \mu^-$})~\cite{Nakamura:2010zzi}, resulting in a value of \mbox{$(2.3 \pm 0.9)\times 10^{-8}$}. The main SM nonresonant \mbox{\Bqmmmm} decay mode is \mbox{$\Bq \rightarrow \mu^+ \mu^- \gamma\,( \to \mu^+ \mu^-)$}, where one muon pair is produced via an electroweak penguin or box diagram and the other via a virtual photon \mbox{[Fig.~\ref{fig:feyns}\,(b)]}. The branching fraction of \mbox{$\Bq \rightarrow \mu^+ \mu^- \gamma\,( \to \mu^+ \mu^-)$} is expected to be less than 10$^{-10}$~\cite{mnt}. The diagram for the MSSM decay mode \BqSP is shown in Fig.~\ref{fig:feyns}\,(c).

 This Letter presents a search for the decays \Bqmmmm using data corresponding to an integrated luminosity of 1.0~\invfb of $pp$ collision data at $\sqrt{s} = 7$\,\tev collected with the LHCb detector in 2011. The resonant \BsJphi decay mode is removed in the signal selection and is used as a control channel to develop the selection criteria. The decay \mbox{\BdJpsiKst}, where \mbox{\Jpsimumu} and \mbox{\KstKpi}, is used as a normalization channel to measure the branching fractions of the  \mbox{ \Bqmmmm} decays.
 
 The detector~\cite{Alves:2008zz} is a single-arm forward
spectrometer, covering the \mbox{pseudorapidity} range $2<\eta <5$, designed
for the study of particles containing $b$ or $c$ quarks. The
detector includes a high precision tracking system consisting of a
silicon-strip vertex detector surrounding the $pp$ interaction region,
a large-area silicon-strip detector located upstream of a dipole
magnet, and three stations
of silicon-strip detectors and straw drift tubes placed
downstream. The magnet has a bending power of about $4{\rm\,Tm}$. The combined tracking system has a momentum resolution
$\Delta p/p$ that varies from 0.4\,\% at momenta of 5\,\gevc to 0.6\,\% at 100\,\gevc. The impact parameter (IP) resolution is 20\,\mum for tracks with high transverse momentum ($p_{\mathrm{T}}$). Charged particles are identified using two
ring-imaging \v{C}erenkov detectors. Photon, electron and hadron
candidates are identified by a calorimeter system consisting of
scintillating-pad and preshower detectors, an electromagnetic
calorimeter and a hadronic calorimeter. Muons are identified by a system composed of alternating layers of iron and multiwire
proportional chambers.  

The trigger consists of a hardware stage, based on information from the calorimeter and muon systems, followed by a software stage that performs a full event reconstruction. Events are selected by the single-muon, dimuon and generic $b$-hadron triggers described in Ref.~\cite{Aaij:2012me}. 
 
Simulated events are generated using \pythia~6.4~\cite{Sjostrand:2006za} configured with the parameters detailed in Ref.~\cite{LHCb-PROC-2011-005}. Final state QED radiative corrections are included using the {\sc{Photos}} package~\cite{Golonka:2005pn}. The \evtgen~\cite{Lange:2001uf} and  \geant~\cite{Agostinelli:2002hh,*Allison:2006ve,1742-6596-331-3-032023} packages are used to generate hadron decays and simulate interactions in the detector, respectively.


Signal \Bqmmmm candidates are selected by applying cuts on the final state muons and the reconstructed \Bq meson. Each final state muon candidate track is required to be matched with hits in the muon system~\cite{muonid}. The muon candidates are required to have particle identification (PID) criteria consistent with those of a muon and not those of a kaon or pion. This is determined by calculating an overall event likelihood for the distribution of \v{C}erenkov photons detected by the ring-imaging \v{C}erenkov system being consistent with a given particle hypothesis. To assess a particle hypotheses the difference between the logarithm of its likelihood and the pion hypothesis likelihood (DLL) is computed. Each muon candidate is required to have \mbox{DLL($K-\pi$)$<0$} and  \mbox{DLL($\mu-\pi$)$>0$}. The PID selection criteria yield a muon selection efficiency of 78.5\,\%; the corresponding efficiency for mis-identifying a pion (kaon) as a muon is 1.4\,\% ($<0.1$\,\%). All muon candidates are required to have a track fit $\
chi^2$ 
per 
degree of freedom of less than 5. Selection criteria are applied on the consistency of the muons to originate from a secondary vertex rather than a primary vertex. The muons are each required to have the difference between the $\chi^2$ of the primary vertex formed with and without the considered tracks, $\chi^2_{\mathrm{IP}}$, to be greater than 16. The \Bq candidates are formed from two pairs of oppositely charged muons. The \Bq decay vertex $\chi^2$ is required to be less than 30 to ensure that the four muons originate from a single vertex. In addition, the reconstructed \Bq meson is required to have a $\chi^2_{\mathrm{IP}} $ less than 9 and hence be consistent with originating from a primary vertex. 

The \Bq candidates are divided into two samples according to the invariant mass of the muon pairs. Signal nonresonant candidates are required to have all four $\mu^+\mu^-$ invariant mass combinations outside the respective $\phi$ and \Jpsi mass windows of $950-1090\,\mevcc$ and $3000-3200\,\mevcc$. The four-muon mass resolution is estimated by the simulation to be around $19\,\mevcc$ for both \Bd and \Bs decay modes. The signal candidates are selected in a four-muon invariant mass window of $\pm40$\,\mevcc around the world average \Bq mass~\cite{Nakamura:2010zzi}. Candidate \BsJphi decays are used to optimize the selection criteria described above. The \BsJphi candidates are selected by requiring the invariant mass of one muon pair to be within the $\phi$ mass window and that of the other pair to be within the \Jpsi mass window. 

The selection criteria are chosen by applying initial cut values that select generic $B$ meson decays. 
These values are then further optimized using \BsJphi candidates. After applying the selection, seven \BsJphi candidates are observed in the signal sample. This is consistent with the expected \mbox{\BsJphi} yield, $5.5 \pm 2.3$, calculated by normalizing to the \BdJpsiKst decay mode.

The dominant \Bqmmmm background is combinatorial, where a candidate \Bq vertex is constructed from four muons that did not originate from a single \Bq meson. Sources of peaking background are estimated to be negligible, the largest of
these is due to \mbox{$\Bd \to \psitwos\,(\to \mu^+ \mu^-)\, K^{*0} \,(\to K^+ \pi^-)$} decays, which has an expected yield of $0.44 \pm 0.06$ events across the entire four-muon invariant mass range of 4776$-$5426\,\mevcc. 

To evaluate the combinatorial background, a single exponential probability distribution function (PDF) is used to fit the events in mass ranges of 4776$-$5220 and 5426$-$5966\,\mevcc, where no signal is expected. Extrapolating the PDF into the \Bd (\Bs) signal window results in an expected background of $0.38^{+0.23}_{-0.17}$ ($0.30^{+ 0.22}_{- 0.20}$) events. Linear and double exponential fit models give consistent background yields.


 The branching fraction of the \mbox{\Bqmmmm} decay is measured relative to that of the normalization channel  \mbox{\BdJpsiKst}, this avoids uncertainties associated with the \Bd production cross section and the integrated luminosity. This normalization channel has the same topology as the signal channel and two muons as final state particles. The $S$-wave component from the nonresonant decay \BdJpsiKpi is removed from the present search. The branching fraction of \mbox{\BdJpsimumuKstKpi} is calculated as the product of \BF(\BdJpsiKst), with the $S$-wave component removed~\cite{Abe:2002haa}, \BF(\KstKpi) and \BF(\Jpsimumu)~\cite{Nakamura:2010zzi}; the resulting branching fraction is  \mbox{$(5.10\pm 0.52)\times10^{-5}$}.

The muon PID and kinematic selection criteria for the normalisation channel are identical to those applied to the signal channel. In addition to these criteria, the kaon is required to have its PID consistent with that of a kaon and not a pion, and vice versa for the pion. This removes \BdJpsiKst candidates where the kaon and pion mass hypotheses are exchanged. The \Kst (\Jpsi) meson is selected by applying a mass window of $\pm 100$\,\mevcc ($\pm 50$\,\mevcc) around the world average invariant mass~\cite{Nakamura:2010zzi}.

There are four main sources of peaking backgrounds for \BdJpsiKst decays. The first is the decay \mbox{$B^+ \rightarrow \Jpsi K^+$} combined with a pion from elsewhere in the event, which is removed by applying a veto on candidates with a $K^+\mu^+\mu^-$ invariant mass within $\pm 60$\,\mevcc of the world average $\Bp$ mass~\cite{Nakamura:2010zzi}. The second arises from $\Bs \rightarrow J/\psi \phi$ decays, where the $\phi$ meson decays to two kaons, one of which is misidentified as a pion. This background is suppressed with a veto on candidates with a $K^+\pi^-$ invariant mass, where the pion is assigned a kaon mass hypothesis, within $\pm70\,\mevcc$ of the world average $\phi$ mass~\cite{Nakamura:2010zzi}. The third source of background is from the decay $\BsbJpsiKst$, which is included in the $K^+ \pi^- \mu^+ \mu^-$ invariant mass fit described below. The last source of background is the $S$-wave component of the decay mode \BdJpsiKpi, this is discussed later.

 \begin{figure}[t]
  \centering
  \includegraphics[width=0.45\textwidth] {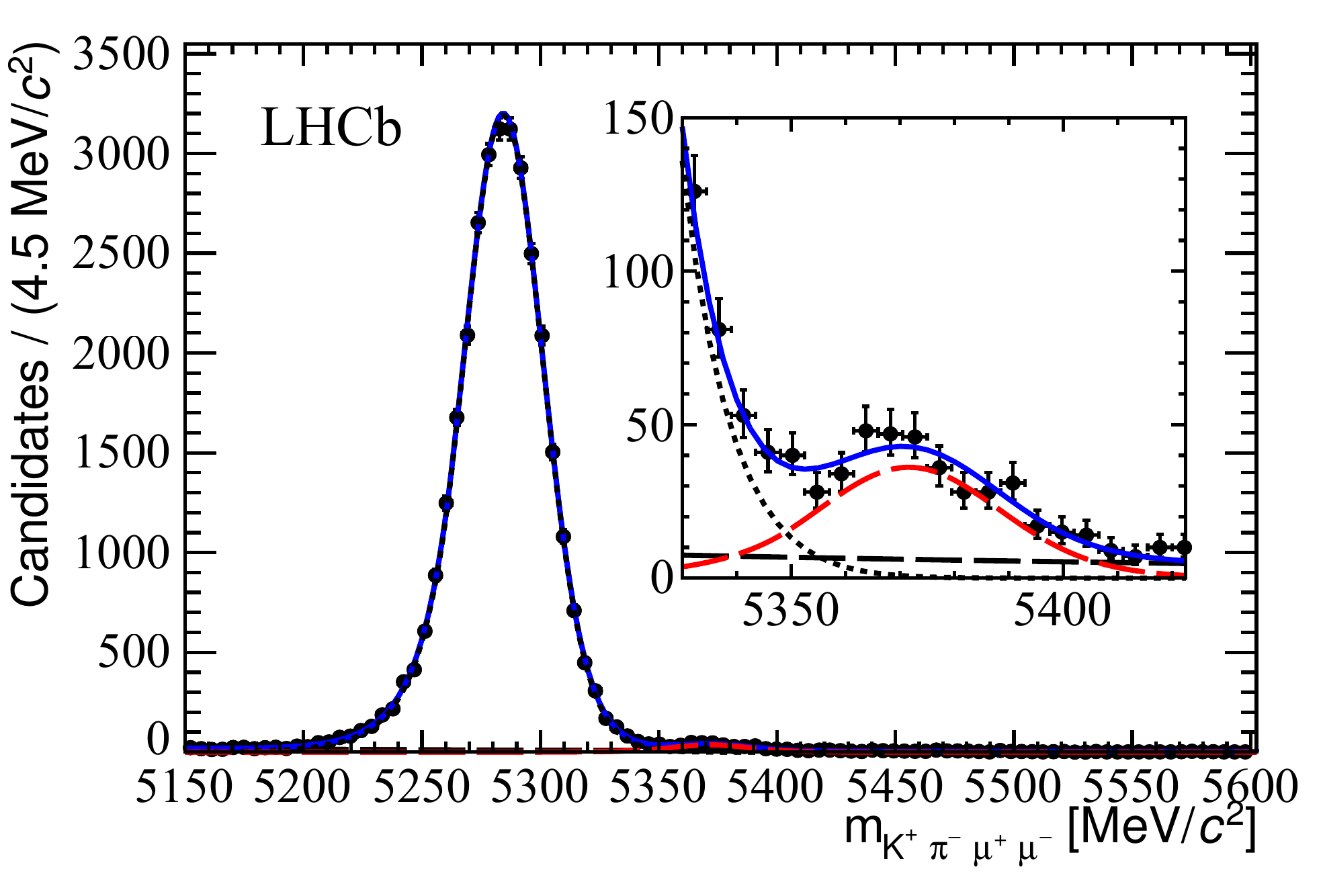}
  \label{fig:jpsikstfit}
  \caption{(color online). Invariant mass distribution of $K^+ \pi^- \mu^+ \mu^-$ candidates. The \Bd and $\Bsb$ signal distributions are shown by short-dashed black and long-dashed red (gray) lines, respectively. The background shape is denoted by a long-dashed black line. The total fit result is shown as a solid blue (black) line. The inset shows the mass distribution centered around the \Bsb mass.}
  \end{figure}

The yield of the normalisation channel is determined by fitting the $K^+ \pi^- \mu^+ \mu^-$ invariant mass distribution with a combination of three PDFs, the \mbox{\BdJpsiKst} and the \mbox{\BsbJpsiKst} signal PDFs consist of the sum of a Gaussian and a Crystal Ball function \cite{Skwarnicki:1986xj}, centred around the respective world average \Bd and \Bs masses and the background PDF consists of a single exponential.  The resulting fit is shown in Fig.~\ref{fig:jpsikstfit}. The \BdJpsiKst mass peak has a Gaussian standard deviation width of $15.9 \pm 0.6$\,\mevcc and contains $31\,800 \pm 200$ candidates.

 The branching fraction of the \Bqmmmm decay is calculated using
 \begin{eqnarray}
  \hspace*{-7mm} \BR(\Bqmmmm) = \BR(\BdJpsiKst)  \times \nonumber\\ 
  &&  \hspace*{-72mm} \frac{\epsilon_{\BdJpsiKst}}{\EBqmmmm}  \frac{\NBqmmmm}{N_{\BdJpsiKst}}\left(\frac{f_{d(s)}}{f_{d}}\right)^{-1}\kappa \text{ ,}
  \label{eq:normalisation}
\end{eqnarray} 
  
  \noindent where \mbox{$\BR(\BdJpsiKst)$} is the branching fraction of the normalisation channel \mbox{\cite{Nakamura:2010zzi,Abe:2002haa}} and $\epsilon_{\BdJpsiKst}$ and $\EBqmmmm$ are the efficiencies for triggering, reconstructing and selecting the normalisation and signal channel events, respectively. The efficiencies are calculated using simulated events and are cross-checked on data. The yields of the normalisation and signal channels are $N_{\BdJpsiKst}$ and $\NBqmmmm$, respectively. The relative production fraction for \Bd and \Bq mesons, $f_{d(s)}/f_d$, is measured to be $f_s/f_d=0.256\pm0.020$ for \Bs decays~\cite{LHCb_fdfs} and taken as unity for \Bd decays. The factor $\kappa$ accounts for the efficiency-corrected $S$-wave contribution to the normalisation channel yield; $\kappa$ is calculated to be $1.09 \pm 0.09$, using the technique described in Ref.~\cite{swave}.

 The PID components of the selection efficiencies are determined from data calibration samples of kaons, pions and muons. The kaon and pion samples are obtained from $D^0\to K^- \pi^+$ decays, where the $D^0$ meson is produced via \mbox{$D^{*+} \to  D^0 \pi^+$} decays. The muon sample is obtained from \mbox{$B^+ \to \Jpsi\,(\to \mu^+ \mu^-)K^+ $} decays. The calibration samples are divided into bins of momentum, pseudorapidity, and the number of charged tracks in the event. This procedure corrects for differences between the kinematic and track multiplicity distributions of the simulated and the calibration event samples.

  Two models are used to simulate \Bqmmmm decays: (i) the phase space model, where the \Bq mass is fixed and the kinematics of the final state muons are distributed according to the available phase-space and (ii) the MSSM model, which describes the decay mode \mbox{\BqSP}. In the MSSM model the pseudoscalar particle $P$ is a sgoldstino of mass $214.3$\,\mevcc, consistent with results from the HyperCP experiment~\cite{Park:2005eka}. The decay widths of $S$ and $P$ are set to $0.1$\,\mevcc.  The scalar sgoldstino $S$ mass is set to 2.5\,\gevcc. If the mass of $S$ is varied across the allowed phase space of the \BqSP decay, the relative change in $\EBqSP$ from the central values varies from $-12.6\,\%$ to $+17.2\,\%$.

\begin{table}
 \caption{Combined reconstruction and selection efficiencies of all the decay modes considered in the analysis. The uncertainties shown are statistical.}

  \begin{tabularx}{0.47\textwidth}{X|X|c}
   
   Decay mode & Model & Efficiency [\%]\Tstrut \\ \hline
   $\Bdmmmm$\Tstrut\Bstrut & Phase space& $0.349 \pm 0.003$ \\ 
   $\BdSP$\Bstrut & MSSM & $0.361 \pm 0.003$ \\ 
   $\Bsmmmm$\Bstrut & Phase space& $0.359 \pm 0.003$ \\ 
   $\BsSP$\Bstrut & MSSM & $0.366 \pm 0.003$ \\ 
   $\BdJpsiKst$ & SM & $0.273 \pm 0.003 $ \\ 
 
\end{tabularx} 

\label{tab:efficiencies}
\end{table}

The calculated efficiencies of all the simulated decay modes are shown in Table \ref{tab:efficiencies}. The total efficiencies of the MSSM models are comparable to those for the phase space models, indicating that the present search has approximately the same sensitivity to new physics models, which feature low mass resonances, as to the phase space models. 

\begin{table}
\caption{Systematic uncertainties on the branching fractions of \Bqmmmm. The combined systematic uncertainties are calculated by adding the individual components in quadrature.}
\centering
 \begin{tabularx}{0.47\textwidth}{lr} 
 
Source & \hspace*{-5.2mm} Systematic uncertainty [\%]\Tstrut \\ \hline
\BR(\BdJpsiKst)\Tstrut &  10.2 \ \ \ \ \ \ \ \ \ \ \ \ \ \ \ \ \\
$S$-wave correction & 8.3 \ \ \ \ \ \ \ \ \ \ \ \ \ \ \ \ \\
$f_d/f_s$ & 7.8 \ \ \ \ \ \ \ \ \ \ \ \ \ \ \ \ \\ 
Data-simulation differences & 5.2 \ \ \ \ \ \ \ \ \ \ \ \ \ \ \ \  \\ 
Trigger efficiency&   4.4 \ \ \ \ \ \ \ \ \ \ \ \ \ \ \ \  \\ 
PID selection efficiency & 4.1 \ \ \ \ \ \ \ \ \ \ \ \ \ \ \ \ \\
Simulation sample size & 1.3 \ \ \ \ \ \ \ \ \ \ \ \ \ \ \ \ \\ 
\BdJpsiKst yield & 0.6 \ \ \ \ \ \ \ \ \ \ \ \ \ \ \ \ \\ \hline
Combined \Bs uncertainty\Tstrut& 17.2 \ \ \ \ \ \ \ \ \ \ \ \ \ \ \ \  \\ 
Combined \Bd uncertainty\Tstrut& 15.4  \ \ \ \ \ \ \ \ \ \ \ \ \ \ \ \ \\ 
 
 \end{tabularx}
\label{tab:combined_sys}

\end{table}

Systematic uncertainties enter into the calculation of the limits on \BR(\Bqmmmm) through the various elements of Eq.~\eqref{eq:normalisation}. The largest uncertainty arises from the branching fraction of \mbox{\BdJpsiKst}, which is known to a precision of 10.2\,\%~\cite{Nakamura:2010zzi,Abe:2002haa}. An uncertainty arises due to the correction for the \BdJpsiKpi $S$-wave contribution. This is conservatively estimated to be 8.3\,\%, which is the maximum relative change in $\kappa$ when it is calculated: (i) by using the angular acceptance from simulated events and (ii) by performing a fit with the physics parameters of the decay fixed and the angular acceptance parameterised, the coefficients of which are left free in the fit. An uncertainty of 7.8\,\% is introduced in the calculation of \mbox{\BR(\Bsmmmm)}, due to the uncertainty on $f_s/f_d$~\cite{LHCb_fdfs}. The 4.4\,\% systematic uncertainty associated 
with the trigger efficiency is calculated as the relative difference between the data and simulation efficiencies of the trigger selection criteria 
applied to the normalisation channel. The efficiency in data is calculated using the method described in Ref.~\cite{TISTOS}. Small differences are seen between the data and the simulated events for the track $\chi^2_{\mathrm{IP}}$ distributions and the efficiency for reconstructing individual tracks. The distributions of these quantities are corrected in the simulation to resemble the data using data-driven methods and the associated uncertainty is assessed by varying the magnitude and the configuration of the corrections. The relative systematic uncertainty assigned to the ratio of efficiencies, $\epsilon_{\BdJpsiKst}/\EBsmmmm$,  is calculated to be 5.2\,\%. The 4.1\,\% uncertainty on the PID selection efficiency is the maximum relative change in $\epsilon_{\BdJpsiKst}/\EBsmmmm$ that results from applying different binning schemes to the PID calibration samples. This uncertainty includes effects associated with muon, kaon and pion identification. The statistical uncertainty associated with the size of the 
simulated event samples is 1.3\,\% for both \Bd and \Bs modes. The uncertainty on the \mbox{\BdJpsiKst} yield is 0.6\,\%. Table~\ref{tab:combined_sys} summarizes the systematic uncertainties for both the \mbox{\BR(\Bsmmmm)} and \mbox{\BR(\Bdmmmm)} branching fractions. The combined systematic uncertainty for the \Bs and \Bd 
modes is 17.2\,\% and 15.4\,\%, respectively. The same uncertainties apply for \mbox{\BR(\BsSP)} and \mbox{\BR(\BdSP)}.


\begin{figure}[t] 
  \centering
  \includegraphics[width=0.45\textwidth] {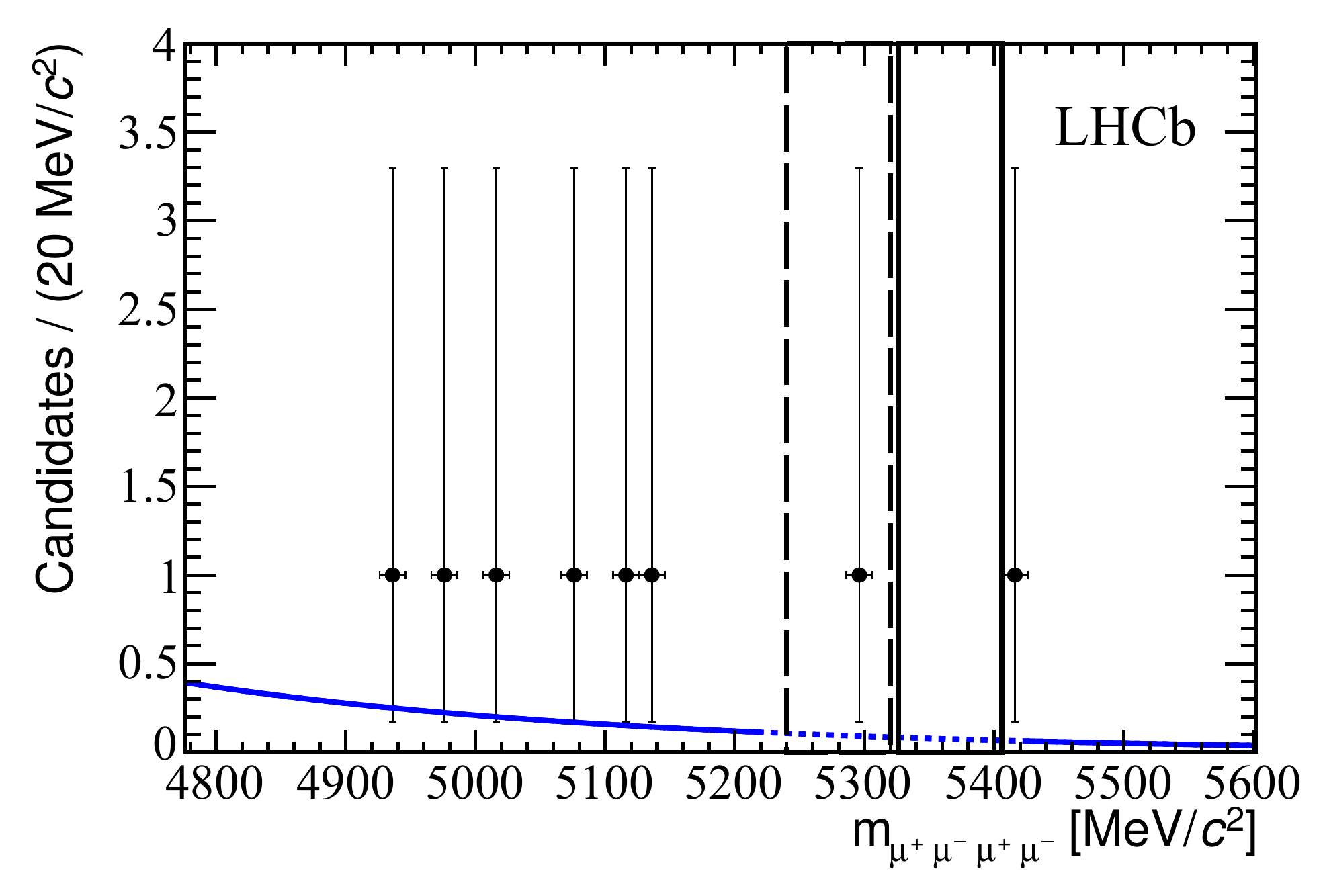}
  \label{fig:unblinded}
  \caption{Invariant mass distribution of nonresonant \Bqmmmm candidates. The solid (dashed) black lines indicate the boundaries of the \Bs (\Bd) signal window. The blue curve shows the model used to fit the mass sidebands and extract the expected number
of combinatorial background events in the \Bs and \Bd signal regions. Only events in the region in
which the line is solid have been considered in the fit.}
  \end{figure} 
  
 The nonresonant four-muon invariant mass range and the background fit are shown in Fig.~\ref{fig:unblinded}. One event is observed in the \Bd signal window and zero events are observed in the \Bs window. These observations are consistent with the expected background yields. The CL$_s$ method ~\cite{Read_02, Junk_99} is used to set upper limits on the branching fractions. The 95\,\% (90\,\%) confidence level limits for the nonresonant \Bqmmmm decay modes in the phase space model are

\begin{equation}
 \BR(\Bsmmmm) < 1.6 \ (1.2) \times 10^{-8} \text{, }\nonumber
\end{equation}
\begin{equation}
 \BR(\Bdmmmm)< 6.6 \ (5.3) \times 10^{-9} \text{.}\nonumber
\end{equation}

 The corresponding limits for the MSSM model with \BqSP and the mass of $P$($S$) set to 214.3\,\mevcc \mbox{(2.5\,\gevcc)}, are
 \begin{equation}
  \BR(\BsSP) < 1.6 \ (1.2) \times 10^{-8}  \text{, }\nonumber
 \end{equation}
 \begin{equation}
   \hspace{-1.5mm} \BR(\BdSP)< 6.3 \ (5.1) \times 10^{-9}\text{.}\nonumber
 \end{equation}

Varying the mass of $S$ across the allowed phase space of the \BqSP decay, from 211\,\mevcc to 5065 (5153)\,\mevcc for \Bd(\Bs),   results in a relative change in the 95\,\% confidence level limit from $-23\,\%$ to $+6\,\%$ for both \Bd and \Bs decay modes.

In summary, a search for the decays \Bqmmmm has been performed and first limits on the branching fractions for these decay modes have been set. These limits probe the upper regions of the parameter space of the \BqSP decay, and, in particular set the first constraints on type-II couplings~\cite{Demidov}.\\

 We express our gratitude to our colleagues in the CERN
accelerator departments for the excellent performance of the LHC. We
thank the technical and administrative staff at the LHCb
institutes. We acknowledge support from CERN and from the national
agencies: CAPES, CNPq, FAPERJ and FINEP (Brazil); NSFC (China);
CNRS/IN2P3 and Region Auvergne (France); BMBF, DFG, HGF and MPG
(Germany); SFI (Ireland); INFN (Italy); FOM and NWO (The Netherlands);
SCSR (Poland); ANCS/IFA (Romania); MinES, Rosatom, RFBR and NRC
``Kurchatov Institute'' (Russia); MinECo, XuntaGal and GENCAT (Spain);
SNSF and SER (Switzerland); NAS Ukraine (Ukraine); STFC (United
Kingdom); NSF (USA). We also acknowledge the support received from the
ERC under FP7. The Tier1 computing centres are supported by IN2P3
(France), KIT and BMBF (Germany), INFN (Italy), NWO and SURF (The
Netherlands), PIC (Spain), GridPP (United Kingdom). We are thankful
for the computing resources put at our disposal by Yandex LLC
(Russia), as well as to the communities behind the multiple open
source software packages that we depend on.

\ifx\mcitethebibliography\mciteundefinedmacro
\PackageError{LHCb.bst}{mciteplus.sty has not been loaded}
{This bibstyle requires the use of the mciteplus package.}\fi
\providecommand{\href}[2]{#2}

\end{document}